\documentstyle[aps,prb,preprint]{revtex}

\begin{document}

\draft

\title
{\bf\Large{Charged and spin-excitation gaps in half-filled strongly
correlated electron systems: A rigorous result}}

\author{Guang-Shan Tian}

\address{Department of Physics, Hong Kong Baptist University,\\
Kowloon Tong, Kowloon, Hong Kong\\
and\\
Department of Physics, Peking University, Beijing 100871, China}

\date{February 6, 1998}

\maketitle

\begin{abstract}

  By exploiting the particle-hole
symmetries of the Hubbard model, the periodic Anderson
model and the Kondo lattice model at half-filling and
applying a generalized version of Lieb's spin-reflection
positivity method, we show that the charged gaps of these
models are always larger than their spin excitation
gaps. This theorem confirms the previous results
derived by either the variational approach or the density
renormalization group approach.

\end{abstract}




  Since the discovery of high temperature superconductivity
in the rare-earthy-based copper oxides \cite{Bednorz}, interest
in the itinerant strongly correlated electron systems has exploded.
The main concern of physicists is the interplay between the itinerant
magnetism and the metallic behavior in these systems. As the typical
models of the strongly correlated electron systems, the Hubbard
model \cite{Hubbard}, the periodic Anderson model \cite{Lee}
and the Kondo lattice model \cite{Aeppli} have been widely studied
in the past several decades.

  To understand the quantum transport and the magnetic properties
of the strongly correlated electron systems, many researcher's interests
have focused on the charged gaps and the spin excitation gaps
in these models at some specific fillings, in particular, at
half-filling. For instance, by solving exactly the one-dimensional
Hubbard model, Lieb and Wu showed that the model has a nonvanishing
charged gap at half-filling for any on-site Coulomb repulsion $U>0$
\cite{Lieb1}. Consequently, in one dimension,
the half-filled Hubbard model is an insulator. This conclusion
is also believed to be true in two dimensions due to the existence
of the spin-wave excitations, which is caused by the nesting Fermi
surface at half-filling, in the system. On the other hand, the spin 
excitation gap of the model is closed \cite{Schrieffer}.

  For the periodic Anderson model, the situation is more
complicated \cite{Nishino}, \cite{Moller}, \cite{Vekic}, \cite{Guerrero}.
For instance, by applying the mean field slave-boson theory,
M\"oller and W\"olfle showed \cite{Moller} that there exists
a critical value $J_c\equiv 4V^2/U \approx 0.036 W$, where
$W$ is the bandwidth and equal to $4dt$ for an $d$-dimensional
simple cubic lattice, such that, when the hybridization energy
$V$ and the on-site Coulomb repulsion between $d$-electrons
$U$ satisfy $4V^2/U<J_c$, the periodic Anderson model at half-filling
has an antiferromagnetic long-range order and the spin excitation
gap vanishes. On the other hand, when $4V^2/U>J_c$, the model has
a nonvanishing spin excitation gap and its ground state is paramagnetic.
The similar conclusions have also been reached by numerical
calculations \cite{Nishino}. However, for the Kondo lattice model,
such a transition point is absent \cite{Tsunetsugu1}, \cite{Yu},
\cite{Shibata}. By using the density-matrix renormalization
group method, Yu and White found that, in one dimension,
the spin excitation gap is nonvanishing for any finite exchange
interaction $J>0$ when the model is half-filled \cite{Yu}.
Their result was further confirmed by Shibata {\it et. al.}
\cite{Shibata}.

  An interesting observation made by these authors is that:
{\it At half-filling, the charged gaps of these models are always
larger than their spin excitation gaps} \cite{Nishino}, \cite{Vekic}
\cite{Yu}. Therefore, an interesting question arose is whether
this observation can be re-established on a more rigorous basis.
In this paper, by using a generalized version of
Lieb's spin-reflection positivity technique \cite{Lieb2},\cite{Lieb3},
\cite {Lieb4}, \cite{Tian1}, we would like to prove
this fact in a mathematically
rigorous way. As a by-product of our proof, one can easily see that
this observation is a result of the particle-hole symmetry
enjoyed by these models at half-filling.

  To begin with, we would first like to introduce several definitions
and some useful notation.

  Take a finite $d$-dimensional simple cubic lattice $\Lambda$
with an {\it even} number of lattice sites and impose the periodic
boundary condition on it. Then, the Hamiltonian of the
Hubbard model can be written as
\begin{eqnarray}
H_{H}
& = &
- t \sum_\sigma \sum_{<{\bf ij}>}
\left(c_{{\bf i}\sigma}^\dagger c_{{\bf j}\sigma} +
c_{{\bf j}\sigma}^\dagger c_{{\bf i}\sigma}\right) \nonumber\\
& + &
U \sum_{{\bf i}\in\Lambda}
\left(n_{{\bf i}\uparrow} - \frac{1}{2}\right)
\left(n_{{\bf i}\downarrow} - \frac{1}{2}\right)
- \mu\hat{N}
\label{Hubbard}
\end{eqnarray}
where $c_{{\bf i}\sigma}^\dagger$ ($c_{{\bf i}\sigma})$
is the fermion creation (annihilation) operator which
creates (annihilates) a fermion with spin $\sigma$ at
lattice site $\bf i$.
$n_{{\bf i}\sigma}=c_{{\bf i}\sigma}^\dagger c_{{\bf i}\sigma}$.
$<{\bf ij}>$ denotes a pair of nearest-neighbor sites
of $\Lambda$. $t > 0$ and $U>0$ are two parameters
representing the kinetic energy and the screened on-site Coulomb
repulsion between fermions, respectively.
$\mu$ is the chemical potential. With respect
to Hamiltonian (\ref{Hubbard}), the simple cubic
lattice is bipartite. In other words, it can be split into two
separate sublattices $A$ and $B$ such that, the fermion hopping
takes place only between sites of different sublattices. In
the following, we shall exploit this fact.

  Similarly, the Hamiltonians of the symmetric periodic Anderson
model and the Kondo lattice model are given by
\begin{eqnarray}
H_A 
& = &
-t \sum_\sigma \sum_{<{\bf ij}>}
\left(c_{{\bf i}\sigma}^\dagger c_{{\bf j}\sigma} +
c_{{\bf j}\sigma}^\dagger c_{{\bf i}\sigma}\right) \nonumber\\
& + &
V \sum_\sigma \sum_{{\bf i}\in\Lambda}
\left(c_{{\bf i}\sigma}^\dagger d_{{\bf i}\sigma} +
d_{{\bf i}\sigma}^\dagger c_{{\bf i}\sigma}\right) \nonumber\\
& + &
U \sum_{{\bf i}\in\Lambda}
\left(d_{{\bf i}\uparrow}^\dagger
d_{{\bf i}\uparrow} - \frac{1}{2}\right)
\left(d_{{\bf i}\downarrow}^\dagger d_{{\bf i}\downarrow}
- \frac{1}{2}\right) - \mu\hat{N}
\label{Anderson}
\end{eqnarray}
\begin{eqnarray}
H_K 
& = & -t \sum_\sigma \sum_{<{\bf ij}>}
\left(c_{{\bf i}\sigma}^\dagger c_{{\bf j}\sigma} +
c_{{\bf j}\sigma}^\dagger c_{{\bf i}\sigma}\right) \nonumber\\
& + &
J \sum_{{\bf i}\in\Lambda}
{\bf\sigma}_{\bf i}\cdot{\bf s}_{\bf i} - \mu\hat{N}
\label{Kondo}
\end{eqnarray}
In Eq.~(\ref{Anderson}), $c_{{\bf i}\sigma}$
and $d_{{\bf i}\sigma}$ represent the atomic $s$-orbital
and $d$-orbital fermion operators at lattice site
$\bf i$, respectively. $V$ stands for the hybridization
energy of $s$-electrons and $d$-electrons.
In Eq.~(\ref{Kondo}), ${\bf\sigma}_{\bf i}$ and ${\bf s}_{\bf i}$
represent the spin operators of the itinerant electrons
and the localized electrons, respectively, and $J>0$ is
the antiferromagnetic exchange interaction between them.

  In terms of Hamiltonians (\ref{Anderson}) and (\ref{Kondo}),
the simple cubic lattice is also bipartite. This fact can be easily
understood by introducing a ``double layer lattice structure''
\cite{Ueda1}, \cite{Tian2}.
For definiteness, let us consider the periodic Anderson model
defined on a specific lattice: The two dimensional square lattice with
the lattice constant being set to be unit. We take two identical copies
of this lattice, $\Lambda_1$ and $\Lambda_2$, and make
a doubly-layered lattice $\widetilde{\Lambda}$ by connecting
the corresponding lattice points of $\Lambda_1$ and $\Lambda_2$
with bonds of length $a=1$. Now, each point of $\widetilde{\Lambda}$
has coordinates ${\bf r} = ({\bf i},\>m)$
with $m = 1, \>2$. Obviously,
$\widetilde{\Lambda}$ has $2N_\Lambda$
lattice points. Next, we define new fermion operators
$e_{{\bf r}\sigma}$ by
\begin{equation}
e_{{\bf r}\sigma} = \left\{
\begin{array}{ll}
c_{{\bf i}\sigma}, & {\rm if} \>\>m = 1;\\
d_{{\bf i}\sigma}, & {\rm if} \>\>m = 2.
\end{array}
\right.
\end{equation}
With the definitions of $\widetilde{\Lambda}$ and $e_{{\bf r}\sigma}$,
the Hamiltonian $H_A$ of the periodic Anderson model can be thought
as the Hamiltonian of a generalized Hubbard model
on the bipartite lattice $\widetilde{\Lambda}$, if $V$ is taken
to be the ``hopping energy'' of $e$-electrons between layer $1$
and layer $2$.

  Similarly, for the Kondo lattice, if we re-define
the partition of sublattices by the following rule: The hopping
energy $t$ and the exchange energy $J$ may be nonzero only
for a pair of sites belonging, respectively, to the different
sublattices, then lattice $\widetilde{\Lambda}$ as well as the original
lattice $\Lambda$ are apparently bipartite
in terms of Hamiltonian (\ref{Kondo}).

  The Hamiltonians $H_H$, $H_A$ and $H_K$ enjoy several symmetries,
which make the analysis of these models easier.

  First, Hamiltonians (\ref{Hubbard}),
(\ref{Anderson}) and (\ref{Kondo})
commute with their total particle number operators
$\hat{N}$, respectively. Consequently, their
Hilbert spaces can be divided into numerous
subspaces $\{V(N)\}$. Each of these subspaces is characterized
by an integer $N$, the total number of fermions in the system.
In particular, the subspace $V(N)$
is called the half-filled subspace if $N=N_\Lambda$
for the Hubbard model, $N=2N_\Lambda$ for both the periodic
Anderson model and the Kondo lattice model. 

  Furthermore, if we define the total spin operators,
for the Hubbard model, by
\begin{eqnarray}
& &
\hat{S}_x \equiv \frac{1}{2} \sum_{{\bf i}\in\Lambda}
\left(c_{{\bf i}\uparrow}^\dagger c_{{\bf i}\downarrow} +
c_{{\bf i}\downarrow}^\dagger c_{{\bf i}\uparrow}\right), \nonumber\\
& &
\hat{S}_y \equiv \frac{1}{2i} \sum_{{\bf i}\in\Lambda}
\left(c_{{\bf i}\uparrow}^\dagger c_{{\bf i}\downarrow} -
c_{{\bf i}\downarrow}^\dagger c_{{\bf i}\uparrow}\right), \nonumber\\
& &
\hat{S}_z \equiv \frac{1}{2} \sum_{{\bf i}\in\Lambda}
\left(n_{{\bf i}\uparrow} - n_{{\bf i}\downarrow}\right)
\end{eqnarray}
for the periodic Anderson model, by
\begin{eqnarray}
& &
\hat{S}_x \equiv \frac{1}{2} \sum_{{\bf i}\in\Lambda}
\left(c_{{\bf i}\uparrow}^\dagger c_{{\bf i}\downarrow} +
c_{{\bf i}\downarrow}^\dagger c_{{\bf i}\uparrow} +
d_{{\bf i}\uparrow}^\dagger d_{{\bf i}\downarrow} +
d_{{\bf i}\downarrow}^\dagger d_{{\bf i}\uparrow}\right),\nonumber\\
& &
\hat{S}_y \equiv \frac{1}{2i} \sum_{{\bf i}\in\Lambda}
\left(c_{{\bf i}\uparrow}^\dagger c_{{\bf i}\downarrow} -
c_{{\bf i}\downarrow}^\dagger c_{{\bf i}\uparrow} +
d_{{\bf i}\uparrow}^\dagger d_{{\bf i}\downarrow} -
d_{{\bf i}\downarrow}^\dagger d_{{\bf i}\uparrow}\right),\nonumber\\
& &
\hat{S}_z \equiv \frac{1}{2} \sum_{{\bf i}\in\Lambda}
\left(n_{{\bf i}\uparrow}^c - n_{{\bf i}\downarrow}^c +
n_{{\bf i}\uparrow}^d - n_{{\bf i}\downarrow}^d\right)
\end{eqnarray}
and, for the Kondo lattice model, by
\begin{eqnarray}
& &
\hat{S}_x \equiv \frac{1}{2} \sum_{{\bf i}\in\Lambda}
\left(c_{{\bf i}\uparrow}^\dagger c_{{\bf i}\downarrow} +
c_{{\bf i}\downarrow}^\dagger c_{{\bf i}\uparrow} +
f_{{\bf i}\uparrow}^\dagger f_{{\bf i}\downarrow}
+ f_{{\bf i}\downarrow}^\dagger f_{{\bf i}\uparrow}\right),\nonumber\\
& &
\hat{S}_y \equiv \frac{1}{2i} \sum_{{\bf i}\in\Lambda}
\left(c_{{\bf i}\uparrow}^\dagger c_{{\bf i}\downarrow} -
c_{{\bf i}\downarrow}^\dagger c_{{\bf i}\uparrow} +
f_{{\bf i}\uparrow}^\dagger f_{{\bf i}\downarrow} -
f_{{\bf i}\downarrow}^\dagger f_{{\bf i}\uparrow}\right),\nonumber\\
& &
\hat{S}_z \equiv \frac{1}{2} \sum_{{\bf i}\in\Lambda}
\left(n_{{\bf i}\uparrow}^c - n_{{\bf i}\downarrow}^c +
n_{{\bf i}\uparrow}^f - n_{{\bf i}\downarrow}^f\right)
\label{Kondo Spin}
\end{eqnarray}
then it is easy to check that the Hamiltonians $H_H$, $H_A$ and $H_K$
commute with their corresponding total spin operators 
$\hat{S}_+=\hat{S}_x+i\hat{S}_y$, $\hat{S}_-=\hat{S}_x-i\hat{S}_y$
and $\hat{S}_z$. Consequently,
any eigenstate $\Psi_n$ with quantum number $S^2=S(S+1)$ of these
Hamiltonians must have $2S+1$ isotopes $\{\Psi_n(M)\}$
with $-S\le M\le S$.

  Here, we would like to emphasize that, in Eq.~(\ref{Kondo Spin}),
$f_{{\bf i}\sigma}^\dagger$ and $f_{{\bf i}\sigma}$ represent
the fermion operators of the localized spins. Consequently, they should
satisfy the following constraint condition \cite{Yanagisawa}
\begin{equation}
f_{{\bf i}\uparrow}^\dagger f_{{\bf i}\uparrow} +
f_{{\bf i}\downarrow}^\dagger f_{{\bf i}\downarrow}
= 1
\label{Constraint}
\end{equation}
This makes them different from operators $c_{{\bf i}\sigma}^\dagger$
and $c_{{\bf i}\sigma}$, which are the itinerant electron operators.

  When the systems are {\it half-filled}, the chemical potential $\mu=0$
for all the Hamiltonians $H_H$, $H_A$ and $H_K$. This fact
is due to the particle-hole symmetry enjoyed by these models
at the special filling \cite{Yang}. As a result, the Hamiltonians
$H_H$, $H_A$ and $H_K$ also commute with the so-called pseudo-spin
operators, which are defined, for the Hubbard model \cite{Yang}, by
\begin{eqnarray}
& &
\hat{J}_+ \equiv \sum_{{\bf i}\in\Lambda} \epsilon({\bf i})
c_{{\bf i}\uparrow}^\dagger c_{{\bf i}\downarrow}^\dagger,\>\>\>\>\>
\hat{J}_- \equiv \hat{J}^\dagger_+, \nonumber\\
& &
\hat{J}_z \equiv \frac{1}{2} \sum_{{\bf i}\in\Lambda}
\left(n_{{\bf i}\uparrow} + n_{{\bf i}\downarrow} - 1\right)
\end{eqnarray}
for the symmetric periodic Anderson model \cite{Nishino}, by
\begin{eqnarray}
& &
\hat{J}_+ \equiv \sum_{{\bf i}\in\Lambda} \epsilon({\bf i})
\left(c_{{\bf i}\uparrow}^\dagger c_{{\bf i}\downarrow}^\dagger -
d_{{\bf i}\uparrow}^\dagger d_{{\bf i}\downarrow}^\dagger\right),
\>\>\>\>\>
\hat{J}_- \equiv \hat{J}^\dagger_+, \nonumber\\
& &
\hat{J}_z \equiv \frac{1}{2} \sum_{{\bf i}\in\Lambda}
\left(n_{{\bf i}\uparrow}^c + n_{{\bf i}\downarrow}^c +
n_{{\bf i}\uparrow}^d + n_{{\bf i}\downarrow}^d - 2\right)
\end{eqnarray}
and, for the Kondo lattice model \cite{Yu}, by
\begin{eqnarray}
& &
\hat{J}_+ \equiv \sum_{{\bf i}\in\Lambda} \epsilon({\bf i})
\left(c_{{\bf i}\uparrow}^\dagger c_{{\bf i}\downarrow}^\dagger -
f_{{\bf i}\uparrow}^\dagger f_{{\bf i}\downarrow}^\dagger\right),
\>\>\>\>\>
\hat{J}_- \equiv \hat{J}^\dagger_+, \nonumber\\
& &
\hat{J}_z \equiv \frac{1}{2} \sum_{{\bf i}\in\Lambda}
\left(n_{{\bf i}\uparrow}^c + n_{{\bf i}\downarrow}^c +
n_{{\bf i}\uparrow}^f + n_{{\bf i}\downarrow}^f - 2\right)
\end{eqnarray}
where $\epsilon({\bf i})=1,\>{\rm for}\>{\bf i}\in A;
\>\epsilon({\bf i})=-1,\>{\rm for}\>{\bf i}\in B$,
is the alternating function. These operators also satisfy
the commutation relations of the conventional spin operators.
Since $H_H$, $H_A$ and $H_K$ commute with $\hat{J}_+$, $\hat{J}_-$
and $\hat{J}_z$, both $J^2$ and $J_z$ are also good quantum numbers
and hence, each eigenstate of $H_H$, $H_A$ and $H_K$ is characterized
by a quantum number $J$ and a quantum number $J_z$ with $-J\le J_z\le J$.
{\it It has been shown that, at half-filling,
the ground states of the Hubbard model \cite{Lieb3},
the symmetric periodic Anderson model \cite{Ueda1} and
the Kondo lattice model \cite{Yanagisawa}, \cite{Tsunetsugu2}
on the simple cubic lattice have quantum numbers $S=0$ and $J=0$}.

  Next, we would like to introduce the definitions of the charged
gaps $\Delta_c$ and the spin excitation gaps $\Delta_s$
for these strongly-correlated electron models at half-filling.
In Refs.~\cite{Nishino} and \cite{Yu}, these quantities are
defined by
\begin{eqnarray}
& &
\Delta_c \equiv E_0(J=1,\>S=0) - E_0(J=0,\>S=0),\nonumber\\
& &
\Delta_s \equiv E_0(J=0,\>S=1) - E_0(J=0,\>S=0)
\end{eqnarray}
where $E_0(J=j,\>S=s)$ is the lowest eigenvalue of the
corresponding Hamiltonian in the subspace with
quantum numbers $J=j$ and $S=s$. By using the definitions
of the pseudo-spin operators and considering the fact that
the ground states of these Hamiltonians at half-filling
have quantum numbers $S=0$ and $J=0$, the above definitions
can be also re-written into the following forms \cite{Shibata},
in terms of the total fermion numbers
\begin{eqnarray}
& &
\Delta_c \equiv E_0(\widetilde{N} + 1) + E_0(\widetilde{N} - 1)
- 2E_0(\widetilde{N}), \nonumber\\
& &
\Delta_s \equiv E_0(\widetilde{N},\>S=1) - E_0(\widetilde{N},\>S=0)
\end{eqnarray}
where $\widetilde{N}=N_\Lambda$ for the Hubbard model and
$\widetilde{N}=2N_\Lambda$ for both the periodic Anderson model
and the Kondo lattice model. Notice that, at half-filling,
the Hamiltonians $H_H$, $H_A$ and $H_K$ enjoy the particle-hole
symmetry and hence, identity
$E_0(\widetilde{N}+1)=E_0(\widetilde{N}-1)$ holds \cite{Yang}.
Therefore, $\Delta_c$ can be further written as
\begin{equation}
\Delta_c = 2 \left[E_0(\widetilde{N}+1) - E_0(\widetilde{N})\right]
\end{equation}

  With these preparations, we shall now summarize our main
result in the following theorem.

  {\bf Theorem:} For the Hamiltonians $H_H$, $H_A$ and $H_K$
defined on a $d$-dimensional simple cubic lattice, when the
system is half-filled, the charged gaps and the corresponding
spin excitation gaps satisfy the following inequality
\begin{equation}
\Delta_c \ge \Delta_s
\label{Inequality}
\end{equation}

 {\it Proof of the theorem:} To prove this theorem, we shall
apply a generalized version of Lieb's spin-reflection positivity
method \cite{Lieb3}, \cite{Lieb4}, which we previously used
to study the binding energy of fermions
in the negative-$U$ Hubbard model \cite{Tian1}. In order to make
it more readable, we shall divide the proof of this theorem
in several steps:
  
  (1) First, by introducing a unitary partial particle-hole
transformation for each model, we map the original Hamiltonians
to some equivalent Hamiltonians with {\it negative} coupling
constants.

  (2) Then, we write each of the transformed Hamiltonians
into a form of the direct product of up-spin fermion operators
and down-spin fermion operators.

  (3) We shall then apply the spin-reflection positivity method
to these transformed Hamiltonians and establish an inequality
for the lowest eigenvalues of these transformed Hamiltonians
in the different subspaces.
 
  (4) Finally, we apply the inverse of the partial particle-hole
transformations to the inequality and finish the proof
of the theorem.

  In the following, to avoid unnecessary digression, we may directly
apply some well-established mathematical results without proving
them. Naturally, in that case, we shall refer to some standard
references for the reader's convenience.

  We now proceed to the proof of the theorem.

  {\it Step 1:} It is a well-known fact that, for the Hamiltonians
$H_H$, $H_A$ and $H_K$ at half-filling, there exist unitary
transformations $\hat{U}_H$, $\hat{U}_A$ and $\hat{U}_K$, which
are called the partial particle-hole transformations \cite{Yang}. Under
these transformations, each Hamiltonian with {\it positive}
interaction coupling constants is mapped to a corresponding
Hamiltonian with {\it negative} interactions. To be more precise,
let us consider these Hamiltonians one by one.

  (a) For the Hubbard Hamiltonian (\ref{Hubbard}), $\hat{U}_H$
is defined by \cite{Lieb2}, \cite{Yang}
\begin{equation}
\hat{U}_H^\dagger c_{{\bf i}\uparrow} \hat{U}_H
= c_{{\bf i}\uparrow},\>\>\>\>\>
\hat{U}_H^\dagger c_{{\bf i}\downarrow} \hat{U}_H
= \epsilon({\bf i}) c_{{\bf i}\downarrow}^\dagger
\end{equation}
At half-filling ($\mu=0$), under $\hat{U}_H$, $H_H$
is mapped to
\begin{eqnarray}
\widetilde{H}_{H}
& = &
\hat{U}_H^\dagger H_H \hat{U}_H =
- t \sum_\sigma \sum_{<{\bf ij}>}
\left(c_{{\bf i}\sigma}^\dagger c_{{\bf j}\sigma} +
c_{{\bf j}\sigma}^\dagger c_{{\bf i}\sigma}\right) \nonumber\\
& - &
U \sum_{{\bf i}\in\Lambda}
\left(n_{{\bf i}\uparrow} - \frac{1}{2}\right)
\left(n_{{\bf i}\downarrow} - \frac{1}{2}\right)
\label{Hubbard2}
\end{eqnarray}
Apparently, $\widetilde{H}_H$ has the same form as $H_H$. However,
in Hamiltonian (\ref{Hubbard2}), the sign of $U$ is changed.

  (b) For the symmetric periodic Anderson model, the unitary transformation
$\hat{U}_A$ is defined by \cite{Ueda1}
\begin{eqnarray}
& &
\hat{U}_A^\dagger c_{{\bf i}\uparrow} \hat{U}_A
= c_{{\bf i}\uparrow},\>\>\>\>\>
\hat{U}_A^\dagger c_{{\bf i}\downarrow} \hat{U}_A
= \epsilon({\bf i}) c_{{\bf i}\downarrow}^\dagger, \nonumber\\
& &
\hat{U}_A^\dagger f_{{\bf i}\uparrow} \hat{U}_A
= f_{{\bf i}\uparrow},\>\>\>\>\>
\hat{U}_A^\dagger f_{{\bf i}\downarrow} \hat{U}_A
= - \epsilon({\bf i}) f_{{\bf i}\downarrow}^\dagger
\end{eqnarray}
Consequently, under $\hat{U}_A$, $H_A$ is mapped to
\begin{eqnarray}
\widetilde{H}_A
& = &
\hat{U}_A^\dagger H_A\hat{U}_A =
- t \sum_\sigma \sum_{<{\bf ij}>}
\left(c_{{\bf i}\sigma}^\dagger c_{{\bf j}\sigma} +
c_{{\bf j}\sigma}^\dagger c_{{\bf i}\sigma}\right) \nonumber\\
& + &
V \sum_\sigma \sum_{{\bf i}\in\Lambda}
\left(c_{{\bf i}\sigma}^\dagger d_{{\bf i}\sigma} +
d_{{\bf i}\sigma}^\dagger c_{{\bf i}\sigma}\right) \nonumber\\
& - &
U \sum_{{\bf i}\in\Lambda}
\left(n_{{\bf i}\uparrow}^d - \frac{1}{2}\right)
\left(n_{{\bf i}\downarrow}^d - \frac{1}{2}\right)
\label{Anderson2}
\end{eqnarray}
Notice that the sign of $U$ is changed but the sign of $V$ is
invariant. Since the hybridization term can be mathematically
treated as a generalized ``hopping'' term, as we shall show,
the sign of $V$ plays no role in the following proof.

  (c) For the Kondo lattice model, the unitary partial
particle-hole transformation $\hat{U}_K$ is given by \cite{Yanagisawa}
\begin{eqnarray}
& &
\hat{U}_K^\dagger c_{{\bf i}\uparrow} \hat{U}_K
= c_{{\bf i}\uparrow},\>\>\>\>\>
\hat{U}_K^\dagger c_{{\bf i}\downarrow} \hat{U}_K
= \epsilon({\bf i}) c_{{\bf i}\downarrow}^\dagger, \nonumber\\
& &
\hat{U}_K^\dagger f_{{\bf i}\uparrow} \hat{U}_K
= f_{{\bf i}\uparrow},\>\>\>\>\>
\hat{U}_K^\dagger f_{{\bf i}\downarrow} \hat{U}_K
= - \epsilon({\bf i}) f_{{\bf i}\downarrow}^\dagger
\end{eqnarray}
Under $\hat{U}_K$, the constraint condition (\ref{Constraint})
now reads
\begin{equation}
f_{{\bf i}\uparrow}^\dagger f_{{\bf i}\uparrow} =
f_{{\bf i}\downarrow}^\dagger f_{{\bf i}\downarrow}
\label{Constraint2}
\end{equation}
and the half-filled Hamiltonian $H_K$ is mapped to \cite{Tsunetsugu2}
\begin{eqnarray}
\widetilde{H}_K
& = &
\hat{U}_K^\dagger H_K\hat{U}_K =
- t \sum_\sigma \sum_{<{\bf ij}>}
\left(c_{{\bf i}\sigma}^\dagger c_{{\bf j}\sigma} +
c_{{\bf j}\sigma}^\dagger c_{{\bf i}\sigma}\right) \nonumber\\
& - &
\frac{J}{4} \sum_\sigma \sum_{{\bf i}\in\Lambda}
\left(c_{{\bf i}\sigma}^\dagger c_{{\bf i}\sigma} +
f_{{\bf i}\sigma}^\dagger f_{{\bf i}\sigma}\right) \nonumber\\
& + &
\frac{J}{2} \sum_{{\bf i}\in\Lambda}
\left(c_{{\bf i}\uparrow}^\dagger c_{{\bf i}\uparrow}
f_{{\bf i}\uparrow}^\dagger f_{{\bf i}\uparrow} +
c_{{\bf i}\downarrow}^\dagger c_{{\bf i}\downarrow}
f_{{\bf i}\downarrow}^\dagger f_{{\bf i}\downarrow}\right) \nonumber\\
& - &
\frac{J}{2} \sum_{{\bf i}\in\Lambda}
\left(c_{{\bf i}\uparrow}^\dagger f_{{\bf i}\uparrow}
c_{{\bf i}\downarrow}^\dagger f_{{\bf i}\downarrow} +
f_{{\bf i}\uparrow}^\dagger c_{{\bf i}\uparrow}
f_{{\bf j}\downarrow}^\dagger c_{{\bf i}\downarrow}\right)
\label{Kondo2}
\end{eqnarray}
We notice that, in Hamiltonian (\ref{Kondo2}), the sign of
the last term is {\it negative}. This fact is the basis of our
proof of the theorem for the Kondo lattice model. The first
three terms can be mathematically treated as generalized
``hopping'' terms. We shall see that their signs do not matter.

  Here, we would like to make some remarks.

  {\bf Remark 1:} Under $\hat{U}_H$,
$\hat{U}_A$ and $\hat{U}_K$, {\it the half-filled subspace
for the corresponding Hamiltonian is mapped into itself}.
In particular, since these transformations are unitary,
the ground states of $H_H$, $H_A$ and $H_K$ in the half-filled
subspace $V(\widetilde{N})$
are mapped onto their counterparts for $\widetilde{H}_H$,
$\widetilde{H}_A$ and $\widetilde{H}_K$ in the same subspace.
However, other subspaces are not invariant under these transformations.

  {\bf Remark 2:} Under the partial particle-hole transformations,
the spin operators $\hat{S}_+$, $\hat{S}_-$ and $\hat{S}_z$
related to each Hamiltonian are mapped onto the corresponding
pseudo-spin operators $\hat{J}_+$, $\hat{J}_-$ and $\hat{J}_z$,
and vice versa. Consequently, under these transformations,
an eigenstate $\Psi(J=j,\>S=s)$ of the original Hamiltonians $H_{H,\>A,\>K}$
is mapped onto an eigenstate $\widetilde{\Psi}(J=s,\>S=j)$
of the transformed Hamiltonians $\widetilde{H}_{H,\>A,\>K}$. In particular,
the ground states of the original Hamiltonians in the sector
with quantum numbers $J=j,\>S=s$ is mapped to the ground states
of the transformed Hamiltonians in the sector with quantum numbers
$J=s,\>S=j$. This observation is important in carrying out our
proof in Step 4.

  {\it Step 2}. Next, we would like to write Hamiltonians
(\ref{Hubbard2}), (\ref{Anderson2}) and (\ref{Kondo2}) into
a form of the direct product of up-spin fermion operators
with down-spin fermion operators. For this purpose,
we shall follow Ref.~\cite{Lieb4} and introduce the following
new fermion operators for each Hamiltonian. We let
\begin{equation}
\hat{C}_{{\bf i}\uparrow} \equiv e_{{\bf i}\uparrow},\>\>\>
\hat{C}_{{\bf i}\downarrow} \equiv
(-1)^{\hat{N}_\uparrow} e_{{\bf i}\downarrow}
\label{Fermion Operators}
\end{equation}
where $e_{{\bf i}\sigma}$ stands for $c_{{\bf i}\sigma}$,
$f_{{\bf i}\sigma}$ and $d_{{\bf i}\sigma}$ appearing in
Eqs.~(\ref{Hubbard2}), (\ref{Anderson2}) and (\ref{Kondo2}),
respectively. $N_\uparrow$ is the total number of fermions
with up-spin in the system. Here, we
would like to emphasize that the new fermion operators
$\{\hat{C}_{{\bf i}\downarrow}\}$, now,
{\it commute with} $\{\hat{C}_{{\bf i}\uparrow}\}$.
Consequently, Hamiltonians (\ref{Hubbard2}), (\ref{Anderson2})
and (\ref{Kondo2}) can be, respectively, re-written as
\begin{eqnarray}
\widetilde{H}_H
& = &
\hat{T}_\uparrow \otimes \hat{I}
+ \hat{I} \otimes \hat{T}_\downarrow \nonumber\\
& - &
U \sum_{{\bf i}\in\Lambda}
\left(\hat{n}_{{\bf i}\uparrow} - \frac{1}{2}\right) \otimes
\left(\hat{n}_{{\bf i}\downarrow} - \frac{1}{2}\right)
\label{Hubbard3}
\end{eqnarray}
\begin{eqnarray}
\widetilde{H}_A
& = &
\hat{T}_\uparrow^\prime \otimes \hat{I}
+ \hat{I} \otimes \hat{T}_\downarrow^\prime \nonumber\\
& - &
U \sum_{{\bf i}\in\Lambda}
\left(\hat{n}_{{\bf i}\uparrow}^d - \frac{1}{2}\right) \otimes
\left(\hat{n}_{{\bf i}\downarrow}^d - \frac{1}{2}\right)
\label{Anderson3}
\end{eqnarray}
and
\begin{eqnarray}
\widetilde{H}_K
& = &
\hat{T}_\uparrow^{\prime\prime} \otimes \hat{I}
+ \hat{I} \otimes \hat{T}_\downarrow^{\prime\prime} \nonumber\\
& - &
\frac{J}{2} \sum_{{\bf i}\in\Lambda}
\left(c_{{\bf i}\uparrow}^\dagger f_{{\bf i}\uparrow} \otimes
c_{{\bf i}\downarrow}^\dagger f_{{\bf i}\downarrow} +
f_{{\bf i}\uparrow}^\dagger c_{{\bf i}\uparrow} \otimes
f_{{\bf j}\downarrow}^\dagger c_{{\bf i}\downarrow}\right)
\label{Kondo3}
\end{eqnarray}
In Eqs.~(\ref{Hubbard3}), (\ref{Anderson3}) and (\ref{Kondo3}),
$\hat{T}_\sigma$, $\hat{T}_\sigma^{\prime}$ and
$\hat{T}_\sigma^{\prime\prime}$ are some Hermitian polynomials
of the fermion operators of spin $\sigma$. $\hat{I}$ is
the identity operator.
Apparently, all the above Hamiltonians can be written in the following
standard form
\begin{eqnarray}
\widetilde{H}
& = &
\hat{G}_\uparrow \otimes \hat{I}
+ \hat{I} \otimes \hat{G}_\downarrow \nonumber\\
& - &
\lambda \sum_{{\bf i}\in\Lambda}
\left(\hat{Q}_{{\bf i}\uparrow} \otimes \hat{Q}_{{\bf i}\downarrow} 
+ \hat{Q}_{{\bf i}\uparrow}^\dagger \otimes
\hat{Q}_{{\bf i}\downarrow}^\dagger \right)
\label{Standard}
\end{eqnarray}
where $\lambda$ is a {\it positive} constant. We notice that
all the operators $\{\hat{Q}_{{\bf i}\sigma}\}$
and $\{\hat{Q}_{{\bf i}\sigma}^\dagger\}$
in Eq.~(\ref{Standard}) are {\it real operators}.
In other words, they are polynomials of $\hat{C}_{{\bf i}\sigma}$
and $\hat{C}_{{\bf i}\sigma}^\dagger$ with {\it real coefficients}.
The fact is of fundamental importance for applying Lieb's
spin-reflection positivity method.
Now, we are able to treat $\widetilde{H}_H$, $\widetilde{H}_A$ and
$\widetilde{H}_K$ simultaneously, by studying the standard
Hamiltonian (\ref{Standard}).

  {\it Step 3:} Now, let us consider the ground state
$\Psi_0(\widetilde{N}+1)$ of $\widetilde{H}$ in the subspace
$V(\widetilde{N}+1)$. Since the spin operators $\hat{S}_+$
and $\hat{S}_-$ commute with $\widetilde{H}$, by applying these
operators an appropriate number of times, we can always transform
$\Psi_0(\widetilde{N}+1)$ into a state satisfying
the condition $N_\uparrow-N_\downarrow=1$. This state
has quantum number $S_z=\frac{1}{2}$. In the following, we shall
exclusively use $\Psi_0(\widetilde{N}+1)$ to denote this state.

  The wave function $\Psi_0(\widetilde{N}+1)$, which has
$\frac{\widetilde{N}}{2}+1$ up-spin fermions and
$\frac{\widetilde{N}}{2}$ down-spin fermions, can be
written as
\begin{equation}
\Psi_0(\widetilde{N}+1) = \sum_{m,\>n} W_{mn} \chi_m^\uparrow
\otimes \chi_n^\downarrow
\label{Wave Function}
\end{equation}
In Eq.~(\ref{Wave Function}), $\chi_k^\sigma$ is a state vector
defined by
\begin{equation}
\chi_k^\sigma \equiv \hat{C}_{{\bf i}_1\sigma}^\dagger
\cdots \hat{C}_{{\bf i}_M\sigma}^\dagger \mid 0\rangle
\end{equation}
where $\left({\bf i}_1,\dots,{\bf i}_M\right)$,
$M=\frac{\widetilde{N}}{2}+1$, for $\sigma=\uparrow$;
$M=\frac{\widetilde{N}}{2}$, for $\sigma=\downarrow$,
denote the positions of fermions with spin $\sigma$.
We would like to emphasize that, when
$\widetilde{H}=\widetilde{H}_K$, the constraint condition
(\ref{Constraint2}) should be taken into consideration to
determine the coefficients $\{W_{mn}\}$ of the ground state
wave function. Apparently, the entire set $\{\chi_k^\sigma\}$
gives a natural basis for $V_\sigma(M)$,
the subspace of $M$ fermions with spin $\sigma$.
However, we should notice that, if we naively choose
${\cal H}_\uparrow=V_\uparrow(\frac{\widetilde{N}}{2}+1)$
and ${\cal H}_\downarrow=V_\downarrow(\frac{\widetilde{N}}{2})$
to be the subspaces for up-spin and down-spin fermions,
then the coefficient matrix ${\cal W}=\left(W_{mn}\right)$
will be an $C_{\widetilde{N}}^{\widetilde{N}/2+1}
\times C_{\widetilde{N}}^{\widetilde{N}/2}$
matrix, {\it which is not a square matrix}.
Mathematically, it is rather difficult to deal with
such a matrix. To avoid this nuisance, we shall define
both ${\cal H}_\uparrow$ and ${\cal H}_\downarrow$
by ${\cal H}_\sigma=V_\sigma(\frac{\widetilde{N}}{2})
\oplus V_\sigma(\frac{\widetilde{N}}{2}+1)$.
Consequently, ${\cal H}_\uparrow$ and ${\cal H}_\downarrow$
have the same dimension and hence, matrix ${\cal W}$ can now
be written as an $D\times D$ {\it square matrix} with
$D=C_{\widetilde{N}}^{\widetilde{N}/2}
+C_{\widetilde{N}}^{\widetilde{N}/2+1}$. Explicitly,
we have
\begin{equation}
{\cal W} = \left(
\begin{array}{lll}
O\>& {\cal M}\\
O\>& O
\end{array}
\right)
\label{Matrix}
\end{equation}
In Eq.~(\ref{Matrix}),  $\cal M$ is an
$C_{\widetilde{N}}^{\widetilde{N}/2+1}
\times C_{\widetilde{N}}^{\widetilde{N}/2}$
nonzero matrix and $O$ represents zero matrices. In terms
of matrix $\cal W$, the normalization of $\Psi_0(\widetilde{N}+1)$
is now given by ${\rm Tr}{\cal W}^\dagger{\cal W}=1$. 

  For such a square matrix, we have
the following polar factorization lemma
in matrix theory.

  {\bf Lemma:} Let $A$ be an {\it arbitrary} (Not necessarily
Hermitian) $n\times n$ matrix.
Then, there are two $n\times n$ {\it unitary}
matrices $U$, $V$ and an $n\times n$
{\it diagonal} semi-positive definite matrix
$H$ such that
\begin{eqnarray}
& &
A=UHV, \nonumber\\
& &
h_{mn}=h_m\delta_{mn}\>\>{\rm and}
\>\>h_m\ge 0,\>\>m=1,\cdots,n.
\end{eqnarray}

  One can find the proof of this lemma in a standard
textbook of matrix theory \cite{Book} or read the appendix
of Ref.~\cite{Tian1}.

  Applying this lemma, we can find two unitary matrices $U$,
$V$ and a diagonal positive semidefinite matrix
$H$, such that ${\cal W}=UHV$. Consequently, $\Psi_0(\widetilde{N}+1)$
can be re-written as
\begin{eqnarray}
\Psi_0(\widetilde{N}+1)
& = &
\sum_{m=1}^D \sum_{n=1}^D
W_{mn} \chi_m^\uparrow \otimes
\chi_n^\downarrow \nonumber\\
& = &
\sum_{m=1}^D \sum_{n=1}^D \left(UHV\right)_{mn}
\chi_m^\uparrow \otimes \chi_n^\downarrow\nonumber\\
& = &
\sum_{l=1}^D
h_l \psi_l^\uparrow \otimes
\phi_l^\downarrow
\label{WF2}
\end{eqnarray}
with
\begin{equation}
\psi_l^\uparrow = \sum_{m=1}^D
U_{ml} \chi_m^\uparrow,\>\>\>\>
\phi_l^\downarrow = \sum_{n=1}^D
V_{ln} \chi_n^\downarrow
\end{equation}
Since $U$ and $V$ are unitary,
$\{\psi_l^\uparrow\}$ and 
$\{\phi_l^\downarrow\}$ are also
orthonormal bases in subspaces
${\cal H}_\uparrow$
and ${\cal H}_\downarrow$, respectively. Furthermore,
since $\Psi_0(\widetilde{N}+1)$ is an eigenvector of
$\hat{N}_\uparrow$, we have
\begin{equation}
\hat{N}_\uparrow \mid\Psi_0(\widetilde{N} + 1)\rangle
= \left(\frac{\widetilde{N}}{2} + 1\right)
\mid\Psi_0(\widetilde{N} + 1)\rangle
\end{equation}
or, equivalently,
\begin{eqnarray}
& &
\sum_{l=1}^D h_l \left[\hat{N}_\uparrow\psi_l^\uparrow\right]
\otimes \phi_l^\downarrow \nonumber\\
& = &
\sum_{l=1}^D h_l \left[\left(\frac{\widetilde{N}}{2} + 1\right)
\psi_l^\uparrow\right] \otimes \phi_l^\downarrow
\label{Eigenstate}
\end{eqnarray}
Taking the inner product of Eq.~(\ref{Eigenstate}) with
$\phi_{l^\prime}^\downarrow$ projects out
\begin{equation}
h_{l^\prime} \left[\hat{N}_\uparrow \psi_{l^\prime}^\uparrow\right] =
h_{l^\prime} \left[\left(\frac{\widetilde{N}}{2} + 1\right)
\psi_{l^\prime}^\uparrow\right]
\end{equation}
Consequently, the corresponding state $\psi_l^\uparrow$ is
an {\it eigenvector} of $\hat{N}_\uparrow$ with eigenvalue
$\widetilde{N}/2+1$, if $h_l\ne 0$. The same conclusion can
also be reached for operator $\hat{N}_\downarrow$ and states
$\{\phi_l^\downarrow\}$.

  For technical reasons, the above conclusions are generally
written in the following weaker form
\begin{eqnarray}
& &
\langle\Psi_0(\widetilde{N}+1)\mid \hat{N}_\uparrow
\mid\Psi_0(\widetilde{N}+1)\rangle \nonumber\\
& = &
\sum_{l=1}^D
h_{l}^2  \langle\psi_{l}\mid \hat{N}
\mid\psi_{l}\rangle = \frac{\widetilde{N}}{2} + 1
\label{Constraint3}
\end{eqnarray}
and
\begin{eqnarray}
& &
\langle\Psi_0(\widetilde{N}+1)\mid \hat{N}_\downarrow
\mid\Psi_0(\widetilde{N}+1)\rangle \nonumber\\
& = &
\sum_{l=1}^D
h_{l}^2  \langle\phi_{l}\mid \hat{N}
\mid\phi_{l}\rangle = \frac{\widetilde{N}}{2}
\label{Constraint4}
\end{eqnarray}
In both Eqs.~(\ref{Constraint3}) and (\ref{Constraint4}),
the spin indices are dropped in the sums, because,
in each equation, only one species of spin is involved.
These equations are particularly useful in proving the strictness
of inequality (\ref{Inequality}) for a {\it finite} system (See
Ref.~\cite{Tian1} for details) and will be used in the following.
  
  In terms of Eq.~(\ref{WF2}), 
the ground state energy of $\widetilde{H}$
in subspace $V(\widetilde{N}+1)$ is now given by
\begin{eqnarray}
& &
E_0(\widetilde{N}+1) \nonumber\\
& = &
\langle\Psi_0(\widetilde{N}+1)\mid \widetilde{H} \mid
\Psi_0(\widetilde{N}+1)\rangle \nonumber\\
& = &
\sum_{l=1}^D h_l^2
\left[\langle\psi_l^\uparrow\mid
\hat{G}_\uparrow
\mid\psi_l^\uparrow\rangle +
\langle\phi_l^\downarrow\mid
\hat{G}_\downarrow
\mid\phi_l^\downarrow\rangle\right] \nonumber\\
& - &
\lambda \sum_{{\bf i}\in\Lambda}
\left(\sum_{l_1,\>l_2=1}^D h_{l_1}
h_{l_2} \langle\psi_{l_2}^\uparrow\mid
\hat{Q}_{{\bf i}\uparrow}\mid\psi_{l_1}^\uparrow\rangle
\langle\phi_{l_2}^\downarrow\mid
\hat{Q}_{{\bf i}\downarrow} 
\mid\phi_{l_1}^\downarrow\rangle\right) \nonumber\\
& - &
\lambda \sum_{{\bf i}\in\Lambda}
\left(\sum_{l_1,\>l_2=1}^D h_{l_1}
h_{l_2} \langle\psi_{l_2}^\uparrow\mid
\hat{Q}_{{\bf i}\uparrow}^\dagger\mid\psi_{l_1}^\uparrow\rangle
\langle\phi_{l_2}^\downarrow\mid
\hat{Q}_{{\bf i}\downarrow}^\dagger 
\mid\phi_{l_1}^\downarrow\rangle\right)
\end{eqnarray}
Applying inequality
$\mid ab\mid\le\frac{1}{2}(\mid a\mid^2+\mid b\mid^2)$
to each term in the triple summations and dropping the spin
indices, we obtain
\begin{eqnarray}
& &
E_0(\widetilde{N}+1) \nonumber\\
& \ge &
\frac{1}{2} \sum_{l=1}^D h_l^2 \left[\langle\psi_l\mid
\hat{G} \mid\psi_l\rangle + \langle\psi_l\mid \hat{G}
\mid\psi_l\rangle \right] \nonumber\\
& + &
\frac{1}{2} \sum_{l=1}^D h_l^2
\left[\langle\phi_l\mid \hat{G} \mid\phi_l\rangle +
\langle\phi_l\mid \hat{G} \mid\phi_l\rangle\right] \nonumber\\
& - &
\frac{\lambda}{2} \sum_{{\bf i}\in\Lambda}
\left(\sum_{l_1,\>l_2=1}^D
h_{l_1} h_{l_2} \langle\psi_{l_2}\mid
\hat{Q}_{\bf i} \mid\psi_{l_1}\rangle
\overline{\langle\psi_{l_2}\mid
\hat{Q}_{\bf i}\mid\psi_{l_1}\rangle}\right) \nonumber\\
& - &
\frac{\lambda}{2} \sum_{{\bf i}\in\Lambda}
\left(\sum_{l_1,\>l_2=1}^D
h_{l_1} h_{l_2} \langle\phi_{l_2}\mid
\hat{Q}_{\bf i} \mid\phi_{l_1}\rangle
\overline{\langle\phi_{l_2}\mid
\hat{Q}_{\bf i} \mid\phi_{l_1}\rangle}\right) \nonumber\\
& - &
\frac{\lambda}{2} \sum_{{\bf i}\in\Lambda}
\left(\sum_{l_1,\>l_2=1}^D
h_{l_1} h_{l_2} \langle\psi_{l_2}\mid
\hat{Q}_{\bf i}^\dagger \mid\psi_{l_1}\rangle
\overline{\langle\psi_{l_2}\mid
\hat{Q}_{\bf i}^\dagger \mid\psi_{l_1}\rangle}\right) \nonumber\\
& - &
\frac{\lambda}{2} \sum_{{\bf i}\in\Lambda}
\left(\sum_{l_1,\>l_2=1}^D
h_{l_1} h_{l_2} \langle\phi_{l_2}\mid
\hat{Q}_{\bf i}^\dagger \mid\phi_{l_1}\rangle
\overline{\langle\phi_{l_2}\mid
\hat{Q}_{\bf i}^\dagger \mid\phi_{l_1}\rangle}\right)
\label{Bound1}
\end{eqnarray}

  Next, we introduce new wave functions
$\Psi_1$ and $\Psi_2$ by
\begin{equation}
\Psi_1 = \sum_{l=1}^D h_l \psi_l^\uparrow
\otimes \bar{\psi}_l^\downarrow,\>\>\>
\Psi_2 = \sum_{l=1}^D h_l \phi_l^\uparrow
\otimes \bar{\phi}_l^\downarrow
\end{equation}
where $\bar{\psi}_l$ and $\bar{\phi}_l$
are the complex conjugates of $\psi_l$ and
$\phi_l$, respectively. Apparently,
we have
\begin{eqnarray}
& &
\langle\Psi_1\mid\Psi_1\rangle =
\langle\Psi_2\mid\Psi_2\rangle \nonumber\\
& = &
\sum_{l=1}^D h_l^2 =
\langle\Psi_0(\widetilde{N} + 1)
\mid\Psi_0(\widetilde{N} + 1)\rangle
= 1
\end{eqnarray}
Since $\hat{G}$ is hermitian and $\{\hat{Q}_{\bf i}\}$
($\{\hat{Q}_{\bf i}^\dagger\}$) are {\it real}, in terms of $\Psi_1$
and $\Psi_2$, inequality (\ref{Bound1}) can be re-written as
\begin{equation}
E_0(\widetilde{N}+1) \ge \frac{1}{2}
\langle\Psi_1\mid \widetilde{H} \mid\Psi_1\rangle +
\frac{1}{2} \langle\Psi_2\mid \widetilde{H} \mid\Psi_2\rangle
\end{equation}
On the other hand, we notice that $\Psi_1$ and $\Psi_2$
are actually wave functions in subspace
$V(\frac{\widetilde{N}}{2}+1,\>\frac{\widetilde{N}}{2}+1)$ and
$V(\frac{\widetilde{N}}{2},\>\frac{\widetilde{N}}{2})$,
respectively. For example, by using the
constraint condition (\ref{Constraint3}), we have
\begin{eqnarray}
& &
\langle\Psi_1\mid \hat{N}_\uparrow \mid\Psi_1\rangle
= \langle\Psi_1\mid \hat{N}_\downarrow \mid\Psi_1\rangle \nonumber\\
& = &
\sum_{l=1}^D h_{l}^2
\langle\psi_{l}\mid \hat{N} \mid\psi_{l}\rangle
= \frac{\widetilde{N}}{2} + 1
\end{eqnarray}
Therefore, by the variational principle, we obtain
\begin{equation}
E_0(\widetilde{N}+1) \ge \frac{1}{2}
E_0(\widetilde{N} + 2) + \frac{1}{2} E_0(\widetilde{N})
\label{Bound2}
\end{equation}

  {\bf Remark 3:} We should notice that, while the construction
of $\Psi_1$ and $\Psi_2$ for both $\widetilde{H}_H$ and
$\widetilde{H}_A$ is straightforward, the job should be done
in a more cautious way as far as $\widetilde{H}_K$ is concerned.
We should show that both $\Psi_1$ and $\Psi_2$ satisfy
the constraint condition (\ref{Constraint2}). That can be achieved
in the following way.

  By reorganizing the rows and columns of the coefficient
matrix $\cal W$ (by a unitary transformation), we can always
write it in a ``block diagonal'' form:
${\cal W}=diag\>({\cal W}_1,\>{\cal W}_2,\cdots)$, where each block
${\cal W}_k$ is a {\it square matrix} and corresponds to
a sector in which the distribution of $f$-fermions is specified,
subject to the condition (\ref{Constraint2}). Naturally,
${\cal W}_k$ has also the form of Eq.~({\ref{Matrix}).
Applying the lemma to each of these submatrices
$\{{\cal W}_k\}$, we can construct unitary matrices
$U$ and $V$ with the following block diagonal form:
\begin{equation}
U = diag \left(U_1,\>U_2,\cdots\right),\>\>\>\>\>
V = diag \left(V_1,\>V_2,\cdots\right)
\end{equation}
and diagonalize the coefficient matrix $\cal W$ by them.
Now, it is easy to see that the wave functions
$\Psi_1$ and $\Psi_2$ constructed according to the above-mentioned
rule satisfy the constraint condition (\ref{Constraint2}).

  {\it Step 4:} Finally, we apply the inverse
of $\hat{U}_H$, $\hat{U}_A$ and $\hat{U}_K$ to map the ``negative
coupling'' Hamiltonians $\widetilde{H}_H$, $\widetilde{H}_A$
and $\widetilde{H}_K$ back onto the original Hamiltonians
$H_H$, $H_A$ and $H_K$. We are mainly interested in the change
of inequality (\ref{Bound2}) under these transformations.
For this purpose, let us consider how each of $E_0(\widetilde{N})$,
$E_0(\widetilde{N}+1)$ and $E_0(\widetilde{N}+2)$ changes
under the unitary inverse transformations.

  First, as we mentioned in Remark 2, $E_0(\widetilde{N})$
is {\it unchanged} under the partial particle-hole transformations.
In other words, it still presents the lowest eigenvalues
of the original Hamiltonians $H_H$, $H_A$ and $H_K$
in the half-filled subspace $V(N=\widetilde{N})$. Interestingly,
$E_0(\widetilde{N}+1)$ is also invariant. This is due to
the fact that the partial particle-hole transformations only change
the particle number of down-spin fermions
from $\frac{\widetilde{N}}{2}$ to
$\widetilde{N}-\frac{\widetilde{N}}{2}=\frac{\widetilde{N}}{2}$
and keep the particle number of up-spin fermions,
$N_\uparrow=\frac{\widetilde{N}}{2}+1$ unchanged. Consequently,
the subspace $V(N_\uparrow=\frac{\widetilde{N}}{2}+1,\>
N_\downarrow=\frac{\widetilde{N}}{2})$ is unitarily mapped into
itself. However, the change of $E_0(\widetilde{N}+2)$
demands a careful consideration.

  It has been previously proven that, in the subspace
$V(\widetilde{N}+2)$, the ground states of the Hamiltonians
$\widetilde{H}_H$ \cite{Lieb2}, \cite{Lieb3},
$\widetilde{H}_A$ \cite{Ueda1} and $\widetilde{H}_K$
\cite{Yanagisawa}, \cite{Tsunetsugu2} have quantum numbers
$S=0$ and $J=1$. Therefore, under the inverse transformations
$\hat{U}_H^{-1}$, $\hat{U}_A^{-1}$ and $\hat{U}_K^{-1}$, these
states are mapped onto the ground states of the {\it original
Hamiltonians} $H_H$, $H_A$ and $H_K$ in the subspace with
quantum numbers $J=0$ and $S=1$, as we discussed above in
Remark 2. In other words, these states are the ground states
of the original Hamiltonians in the half-filled
sector with $S=1$. Consequently, we obtain
\begin{equation}
E_0(\widetilde{N}+2;\>\widetilde{H}_{H,A,K}) =
E_0(\widetilde{N},\>S=1;\>H_{H,A,K})
\end{equation}
Therefore, for the Hamiltonians $H_H$, $H_A$ and $H_K$,
inequality (\ref{Bound2}) now reads
\begin{equation}
E_0(\widetilde{N}+1) \ge \frac{1}{2}
E_0(\widetilde{N},\>S=1) + \frac{1}{2} E_0(\widetilde{N},\>S=0)
\label{Bound3}
\end{equation}
or, equivalently
\begin{eqnarray}
& &
E_0(\widetilde{N}+1) - E_0(\widetilde{N}) \nonumber\\
& \ge &
\frac{1}{2} E_0(\widetilde{N},\>S=1)
- \frac{1}{2} E_0(\widetilde{N},\>S=0)
\label{Bound4}
\end{eqnarray}
Multiplying both sides of inequality (\ref{Bound4}) by $2$,
we obtain inequality (\ref{Inequality}). That ends our proof
of the theorem. {\bf QED}.

  Some remarks are in order.
  
  {\bf Remark 4:} Actually, inequality (\ref{Bound4}) can
be slightly strengthened for the Hubbard model and the symmetric
periodic Anderson model at half-filling. Following Ref.~\cite{Tian1},
one can easily show that, for any {\it finite} lattice $\Lambda$,
inequality (\ref{Bound4}) is {\it strict}. In other words,
the systems have a Mott charged gap. However, in the thermodynamic
limit, the inequality resumes the form of Eq.~(\ref{Bound4}).
Therefore, to study the Mott metal-insulator transition of these
models in the thermodynamic limit, one needs to introduce some
more sophisticated method.

  {\bf Remark 5:} In the above proof, we assumed that the coupling
constants in each Hamiltonian are site-independent. That is only
a harmless assumption to make our proof simpler. Actually,
inequalities (\ref{Bound2}) as well as (\ref{Bound4})
still hold for the models with {\it site-dependent} coupling
constants and its proof is essentially unchanged. One can find
more details in Ref.~\cite{Tian1}.

  In summary, in this article, by exploiting the partial
particle-hole symmetry of the Hubbard model, the periodic Anderson
model and the Kondo lattice model at half-filling and
applying a generalized version of Lieb's spin-reflection
positivity method, we proved that the charged gaps of these
models are always larger than their spin excitation
gaps. This theorem confirms the previous results
derived by either the variational approach or the density
renormalization group approach.

\begin{center}
ACKNOWLEDGMENTS
\end{center}
  
  I would like to thank Dr. Lei-Han Tang for very stimulating
discussions. I would also like to thank the Croucher Foundation
for support and the physics
department at Hong Kong Baptist University for their
hospitality. This work was also partially supported by the
Chinese National Science Foundation under grant No. 19574002.

\end{document}